\documentclass[sigconf, authorversion]{acmart}

\usepackage[inline]{enumitem}
\setlist[description]{leftmargin=0pt, labelindent=0pt}

\usepackage[english]{babel}
\usepackage{booktabs}
\usepackage{microtype}
\usepackage{numprint}
\usepackage{tabularx}
\usepackage{units}

\let\oldsection\section
\renewcommand{\section}{\vspace{-0.30\baselineskip}\oldsection}

\let\oldsubsection\subsection
\renewcommand{\subsection}{\vspace{-0.30\baselineskip}\oldsubsection}

\let\oldsubsubsection\subsubsection
\renewcommand{\subsubsection}{\vspace{-0.30\baselineskip}\oldsubsubsection}

\npdecimalsign{.}

\copyrightyear{2020}
\acmYear{2020}
\setcopyright{acmcopyright}\acmConference[SIGIR '20]{Proceedings of the 43rd International ACM SIGIR Conference on Research and Development in Information Retrieval}{July 25--30, 2020}{Virtual Event, China}
\acmBooktitle{Proceedings of the 43rd International ACM SIGIR Conference on Research and Development in Information Retrieval (SIGIR '20), July 25--30, 2020, Virtual Event, China}
\acmPrice{15.00}
\acmDOI{10.1145/3397271.3401298}
\acmISBN{978-1-4503-8016-4/20/07}

\begin{document}

\fancyhead{}

\title{Predicting Entity Popularity to Improve Spoken Entity Recognition by Virtual Assistants}

\author{Christophe Van Gysel}
\email{cvangysel@apple.com}
\orcid{0000-0003-3433-7317}
\affiliation{%
  \institution{Apple}
}

\author{Manos Tsagkias}
\email{etsagkias@apple.com}
\affiliation{%
  \institution{Apple}
}

\author{Ernest Pusateri}
\email{epusateri@apple.com}
\affiliation{%
  \institution{Apple}
}

\author{Ilya Oparin}
\email{ioparin@apple.com}
\affiliation{%
  \institution{Apple}
}

\renewcommand{\shortauthors}{Van Gysel, et al.}

\newcommand{\StartToken}{\text{<s>}}
\newcommand{\EndToken}{\text{</s>}}
\newcommand{\EntityToken}{\textsc{entity}}

\newcommand{\Apply}[2]{#1\left(#2\right)}
\newcommand{\Prob}[2][P]{\Apply{#1}{#2}}
\newcommand{\CondProb}[3][P]{\Prob[#1]{#2 \mid #3}}

\newcommand{\AP}{AP}
\newcommand{\VerboseAP}{average precision}

\newcommand{\VerboseRecallCut}[1]{recall@$#1$}
\newcommand{\VerbosePrecisionCut}[1]{precision@$#1$}

\newcommand{\VerboseWER}{word error rate}
\newcommand{\WER}{WER}
\newcommand{\WERCut}[1]{\WER{}@$#1$}

\begin{abstract}
We focus on improving the effectiveness of a Virtual Assistant (VA) in recognizing emerging entities in spoken queries. We introduce a method that uses historical user interactions to forecast which entities will gain in popularity and become trending, and it subsequently integrates the predictions within the Automated Speech Recognition (ASR) component of the VA. Experiments show that our proposed approach results in a 20\% relative reduction in errors on emerging entity name utterances without degrading the overall recognition quality of the system.
\end{abstract}

\keywords{virtual assistants; spoken IR; entity popularity; ASR}

\maketitle

\section{Introduction}
\label{section:introduction}

Virtual assistants are becoming increasingly popular \citep{Juniper2019popularity} to help users accomplish a variety of common tasks \citep{Graus2016taskreminders,Maarek2019alexa}.
\citet{Maarek2019alexa} points out that the Automated Speech Recognition (ASR) engine~--~virtual assistants' primary component to analyze spoken commands~--~is key to customer experience, as ASR errors have a detrimental effect on the ability of the virtual assistant to respond accurately to user queries. A prevalent and particularly challenging type of query is entity name queries~\citep{Yin2010nameentityqueries}~--~queries that consist of solely an entity's name (e.g., \emph{"Jennifer Aniston"} without any other search terms).
An analysis of a month of anonymized query logs from a popular virtual assistant shows that, for popular entities occurring in knowledge queries (i.e., the 20\% most frequent entities that make up 91\% of knowledge query traffic), the entity's name by itself is the most frequent query form for 45\% of these popular entities. Hence, entity name queries are a frequent use-case of voice search~\citep{Guy2018characteristics}.

ASR systems are reasonably effective for popular entities as there is ample training data \citep{Serrino2019contextualrecovery}, however, effectiveness degrades for less popular entities.
From a customer experience viewpoint, the problem is exacerbated for emerging entities that appear in the spotlight of attention, e.g., a news entity.
In fact, if we consider entities whose query frequency (temporarily) increased by a factor of 3 from June to July 2019, the recognition accuracy of these emerging entities is 20\% worse when compared to popular entities of the same months.

In this paper we focus on improving speech recognition accuracy for name entity queries, i.e., queries that consist of the name of an entity without any context~--~which typically helps recognition; e.g., \emph{"what is borrelia"} gets recognized correctly, while saying \emph{"borrelia"} gets misrecognized as \emph{gorilla}. In particular, we are interested in improving queries revolving around long-tail and emerging entities.
We introduce a hybrid system that brings Information Retrieval (IR) methods into ASR models. Within the ASR part, we introduce a model-agnostic approach that allows tuning the recognition likelihood of a set of given texts at runtime.
We then adapt IR techniques from named entity recognition and entity linking to predict what emerging entities will become trending, i.e., entities that will be increasingly queried in our query logs. We use the outcome of our predictions to influence the output of the ASR system at runtime.

Our research questions are as follows: %
\begin{enumerate*}[label=(\textbf{RQ\arabic*})]
    \item Can we predict entity name popularity in virtual assistant query logs and subsequently use those predictions to improve speech recognition in entity name queries?
    \item Does adding more historical data improve the accuracy of entity name popularity prediction?
    \item Do the various signals (i.e., features) perform well when used in isolation?
\end{enumerate*}
We contribute: %
\begin{enumerate*}[label=(\arabic*)]
    \item a method that improves automatic speech recognition for emerging entity names;
    \item a novel connection between ASR and IR that demonstrates how IR research can be applied to ASR problems, and
    \item an analysis of the signals within the virtual assistant query log that indicates whether entities are becoming increasingly popular.
\end{enumerate*}

\section{ASR primer}
\label{section:methodology:asr_primer}

Automated speech recognition is the task of translating a speech signal $X$ into a string of words $s$. It works by optimizing \citep[p.~289]{Jurafsky2008slp}:
\begin{equation}
\label{eq:asr}
\begin{split}
S^* & = \text{argmax}_s \CondProb{s}{X} = \text{argmax}_s \frac{\CondProb{X}{s}}{\Prob{X}} \Prob{s} \\
    & = \text{argmax}_s \CondProb{X}{s} \Prob{s}, \\
\end{split}
\end{equation}
\clearpage%
\noindent%
where $\CondProb{X}{s}$ is provided by the acoustic model and denotes the likelihood of speech signal $X$ given the string of words $s$, and $\Prob{s}$ is provided by the language model (LM) and denotes the prior probability of a string of words. $\Prob{X}$ is the probability of the speech signal and can be ignored as it is constant for all hypotheses.

The acoustic model provides probability estimates for a sequence of short audio frames (e.g., 10ms) given an input sequence of sound units (phones), and the lexicon that defines possible pronunciations for different words. The LM \citep{jelinek1998,bengio2003} provides probability estimates for words following each other, thus taking lexical context into account.
The acoustic model and lexicon (\S\ref{section:setup:asr}) are out of scope for this paper as we are specifically targetting entity names that were already recognized previously but fail to be recognized as stand-alone queries, and hence, fall into the domain of the LM.

The LM (\S\ref{section:setup:lm}) builds on the chain rule of probability:
\begin{equation}
    \Prob{W} = \Prob{w_1 w_2\dots w_n} = \prod^{N}_{i=1}\CondProb{w_i}{w_1 w_2\dots w_{i-1}}.
\end{equation}
In practice, strings of words are wrapped in special start/end markers ($\StartToken{}$/$\EndToken{}$, resp.) to denote the beginning and the end of the string of words (which is, typically, a sentence). For example, the prior probability of utterance \emph{SIGIR} would be computed as
\begin{equation*}
\Prob{\text{<s> SIGIR </s>}} = \CondProb{\text{SIGIR}}{\StartToken{}} \CondProb{\EndToken{}}{\text{<s> SIGIR}},
\end{equation*}
where <s> and </s> mark the beginning and end of sentence, resp.

\section{Boosting ASR of Entities}
\label{section:methodology:asr_boosting}

Similar to search \citep{Schurman2009latency}, speech systems must return a hypothesis quickly as users expect fast respond times. Hence, Eq.~\ref{eq:asr} is solved approximately using a greedy beam search. In addition, various components of the ASR system can be inexact. Consequently, to improve the recognition of entity names, we increase the probability of the LM to emit an entity such that the likelihood that the correct hypothesis is chosen increases.
Regardless of the exact implementation of the LM conditional probability, $\CondProb{\text{next}}{\text{history}}$, we can artificially enrich the LM training data to include $\text{<s> \EntityToken{} </s>}$ with a predefined probability of $\alpha$, where $\EntityToken{}$ corresponds to a special word which can be substituted dynamically at runtime. More specifically, at runtime, whenever $\EntityToken{}$ occurs within the LM provided by $P$, we insert the conditional probability distribution $\CondProb[Q]{\;\cdot}{\EntityToken{}}$, which equals the predicted popularity of an entity name between the time the ASR system was built and the time it was used to handle user requests. %
In this paper, we select a list of potentially-popular entity names ($\cdot$; out of a large candidate set) and construct $\CondProb[Q]{\;\cdot}{\EntityToken{}}$ by assigning an equal probability to every entity name in the list.

\section{Forecasting trending entity names}
\label{section:methodology:forecasting}

\newcommand{\Entities}{E}
\newcommand{\Entity}[1]{\MakeLowercase{\Entities{}}_{#1}}

\newcommand{\FrequencyFn}{f}

\newcommand{\Frequency}[2]{\Apply{\FrequencyFn{}}{\TimePeriod{#1}, \Entity{#2}}}
\newcommand{\RelativeFrequency}[2]{\CondProb{\Entity{#2}}{\TimePeriod{#1}}}

\newcommand{\TrendingFactor}{c}

\newcommand{\TimeSeriesLength}{n}
\newcommand{\Timestamp}{t}
\newcommand{\VerboseTimePeriod}[2]{\left[\Timestamp{}_{#1}, \Timestamp{}_{#2} \right)}
\newcommand{\TimePeriod}[1]{T_{#1}}
\newcommand{\TimePeriodLength}{\Size{T}}

\newcommand{\TimePeriodEntities}[1]{\Entities{}^{\TimePeriod{#1}}}
\newcommand{\TrendingTimePeriodEntities}[1]{\Apply{\mathbb{T}}{\TimePeriodEntities{#1}}}

\newcommand{\VerboseCandidateEntities}[2]{\bigcup\nolimits_{k = #1}^{#2} \TimePeriodEntities{k}}
\newcommand{\CandidateEntities}[2]{\Entities{}^{\TimePeriod{#2}}_{\TimePeriod{#1}}}

\newcommand{\Size}[1]{\left|#1\right|}

We present a framework for obtaining a set of entity names that are likely to be queried more frequently. Denote $\TimePeriod{i} = \VerboseTimePeriod{i}{i + 1}$ as the time period that starts at time $\Timestamp{}_i$ and ends at time $\Timestamp{}_{i + 1}$ (excl.). We gather entity query (\S\ref{section:setup:named_entities}) statistics over a series of $\TimeSeriesLength{}$ consecutive, equi-length time periods, $\TimePeriod{1}, \ldots, \TimePeriod{\TimeSeriesLength{}}$, each of length $\TimePeriodLength{}$ (1 week in this paper). For every time period $\TimePeriod{i}$, we have a set of entities $\Entity{j} \in \TimePeriodEntities{i}$ with associated query frequency $\Frequency{i}{j}$, i.e., the number of times entity $\Entity{j}$ occurred in a query during time period $\TimePeriod{i}$. The goal is now to predict the set of entities $\TrendingTimePeriodEntities{n} \subset \CandidateEntities{1}{n - 1}$, where $\CandidateEntities{1}{n - 1} = \VerboseCandidateEntities{1}{n - 1}$, whose frequency will increase by a factor of $\TrendingFactor{}$ from time period $\TimePeriod{n - 1}$ to time period $\TimePeriod{n}$, i.e.,
\begin{equation}
\label{eq:trending}
\TrendingTimePeriodEntities{n} = \big\{ \Entity{j} \in \CandidateEntities{1}{n - 1} \mid \Frequency{n}{j} \geq \TrendingFactor{} \cdot \Frequency{n - 1}{j} \big\}.
\end{equation}

\newcommand{\Feature}[1]{F_{#1}}%

\begin{table}
\caption{Overview of time series features used in our machine-learned models. To avoid numerical issues due to extreme values, we define $\log\left(0\right) = 10^{-6}$ and clip the individual features to a maximum value of $10^3$.\label{tbl:features}}
\renewcommand{\arraystretch}{1.0}%
\begin{tabularx}{\columnwidth}{lll}%
\toprule%
& \textbf{Description} & \textbf{Mathematical expression} \\
\midrule
$\Feature{1}$ & \small rel. freq. & \small $\RelativeFrequency{i}{j} = \nicefrac{\Frequency{i}{j}}{\sum_k \Frequency{i}{k}}$ \\
$\Feature{2}$ & \small delta rel. freq. & \small $\RelativeFrequency{i}{j} - \RelativeFrequency{i - 1}{j}$ \\
$\Feature{3}$ & \small delta log rel. freq. & \small $\log\RelativeFrequency{i}{j} - \log\RelativeFrequency{i - 1}{j}$ \\
$\Feature{4}$ & \small rel. delta freq. & \small $\nicefrac{\left( \Frequency{i}{j} - \Frequency{i - 1}{j} \right)}{\Frequency{i - 1}{j}}$ \\
$\Feature{5}$ & \small rel. delta log freq. & \small $\nicefrac{\left( \log\Frequency{i}{j} - \log\Frequency{i - 1}{j} \right)}{\log\Frequency{i - 1}{j}}$ \\
$\Feature{6}$ & \small rel. freq w/ compl. & \small $\RelativeFrequency{i}{j} \cdot \left(1.0 - \RelativeFrequency{i - 1}{j} \right)$ \\
$\Feature{7}$ & \small ratio freq. & \small $\nicefrac{\Frequency{i}{j}}{\Frequency{i - 1}{j}}$ \\
\bottomrule%
\end{tabularx}%
\end{table}

We cast our task as a supervised classification problem where features are extracted from $\TimePeriod{1}, \ldots, \TimePeriod{n - 1}$ and the binary target labels are determined by membership in $\TrendingTimePeriodEntities{n}$. Table~\ref{tbl:features} shows an overview of our features. The number of features varies with the number of weeks we use for feature generation (i.e., $n - 1$). Specifically, if we have $n - 1$ weeks for feature generation, then the number of features equals $\left( n - 1 \right) + \left( n - 2 \right) \cdot 6$ (i.e., one feature per week and 6 features for every pair of consecutive weeks).

\section{Experimental setup}

\newcommand{\BaselineNone}{\textsc{Without boosting}}
\newcommand{\BaselineRandom}{\textsc{Random}}
\newcommand{\BaselinePopularLastWeek}{\textsc{Popular last week}}
\newcommand{\BaselinePopularLastWeekPreviouslyUnpopular}{\textsc{Suddenly popular}}
\newcommand{\BaselineTrendingLastWeek}{\textsc{Trending last week}}
\newcommand{\Adaboost}{\textsc{Adaboost}}
\newcommand{\NN}{\textsc{NN}}

\subsection{Virtual assistant system}
\label{section:setup:asr}
\label{section:setup:tts}
\label{section:setup:lm}
\label{section:setup:named_entities}

We experiment with an ASR system consisting of statically interpolated 4-gram LMs \citep{Church1991gt,Pusateri2019interpolation}, the interpolated LM is entropy pruned \citep{Stolcke2000entropy}. The LMs are trained on a wide variety of data sources, such as manually and automatically transcribed virtual assistant requests and, artificially-generated sources generated from knowledge base statistics. The acoustic model is a convolutional 28M neural network trained from millions of manually transcribed, anonymized virtual assistant requests. The input to the acoustic model consists of 40 mel-spaced filter bank outputs \citep{Jurafsky2008slp}; each frame is concatenated with the preceding and succeeding ten frames, giving an input of 840 dimensions. The lexicon is manually curated by a team of experts and updated regularly with pronunciations of entities.

We set $\alpha$ (\S\ref{section:methodology:asr_boosting}) empirically by analyzing anonymized query logs from January to May 2019; and setting $\alpha$ proportional to the number of times the full query matched a common person name.
When we construct the conditional probability distribution $\CondProb[Q]{\;\cdot}{\EntityToken{}}$ for which entities to boost (\S\ref{section:methodology:asr_boosting}), we filter out entities that can already be correctly recognized by performing a simulation (i.e., a feedback loop) that consists of
\begin{enumerate*}[label=(\alph*)]
    \item automatically synthesizing voice queries for the entire list of entities using a state-of-the-art Text-To-Speech (TTS) system,
    \item performing recognition using the ASR system without boosting entities, and
    \item removing those entities from our list of entities for which the recognition hypothesis is an exact match of the entity name.
\end{enumerate*}
After ASR, voice assistant queries are classified into a set of intents by an LSTM model that uses word embedding features. We focus only on knowledge queries of encyclopedic nature that consist of an entity name with optional context \citep{Reinanda2015entityaspects}. Knowledge queries are parsed by matching entity names in a manually-curated knowledge base that is a superset of Wikipedia.

\subsection{Methods under comparison}

\subsubsection{Machine learning methods}
\label{section:setup:ml}

We experiment with two classifiers (\S\ref{section:methodology:forecasting}):
\begin{enumerate*}[label=(\alph*)]
    \item AdaBoost \citep{Freund1997adaboost} with \numprint{50} decision trees of depth \numprint{1}, and
    \item a feed-forward neural network with \numprint{2} hidden layers, \numprint{128} neurons (Glorot initialization \citep{Glorot2010initialization}) and ReLU \citep{Glorot2011relu} as activation function, trained by minimizing the cross-entropy loss of the logistic activation at the output layer using Adam ($\alpha = 10^{-3}$, $\beta_1 = 0.9$, $\beta_2 = 0.999$, $\epsilon = 10^{-8}$) \citep{Kingma2014adam} for as many iterations until the loss converges.
\end{enumerate*}

\subsubsection{Baselines}
\label{section:setup:baselines}

We consider five baseline systems: an ASR system without boosting entities, and four naive scoring techniques that we use to obtain a ranking of hypotheses: 
\begin{enumerate*}[label=(\alph*)]
    \item \BaselineRandom{}, a random score for every entity,
    \item \BaselinePopularLastWeek{}, the previous week's frequency (Table~\ref{tbl:features}; $\Feature{1}$),
    \item \BaselinePopularLastWeekPreviouslyUnpopular{} (Table~\ref{tbl:features}; $\Feature{6}$), and
    \item \BaselineTrendingLastWeek{}, the ratio of the frequency of the previous week with the week before that (Table~\ref{tbl:features}; $\Feature{7}$).
\end{enumerate*}

\subsection{Datasets and evaluation measures}
\label{section:setup:data}
\label{section:setup:evaluation}

\newcommand{\Significant}{$^*$}%

We use a sub-sample of knowledge-oriented anonymized query logs of a popular voice assistant service that originated from the USA in English between June 10 and August 4, 2019. Sampling from the logs imposes an implicit absolute threshold on the frequency counts, as counts lower than the inverse sample rate are likely to be filtered out.\footnote{We are unable to provide the exact sample rate due to confidentiality; however, the sample rate is less than 15\%.} The sample contains over 300k unique entity names (\S\ref{section:setup:named_entities}). The threshold factor, $\TrendingFactor{}$ (Eq.~\ref{eq:trending}), is set to \numprint{3}. The last week of this period (July 29 to August 4) is used for evaluation (i.e., test set) where the set of trending entities is determined using Eq.~\ref{eq:trending} (\numprint{5583} out of \numprint{126399} entities were trending). The preceding week, July 22--28, is used as target for training models (\numprint{5198} out of \numprint{125934} entities were trending). For feature generation (i.e., the input to our models; \S\ref{section:methodology:forecasting}), unless otherwise stated, we use three weeks of statistics (i.e., features are generated from July 1--21 for training and July 8--28 for testing) within our experiments.

We measure the effectiveness of our approach by evaluating both the classification and ASR accuracy on the test set. More specifically, we measure:
\begin{enumerate*}[label=(\alph*)]
    \item the classification effectiveness, by computing \VerboseAP{} (\AP{}), \VerboseRecallCut{k}, \VerbosePrecisionCut{k} for $k = 2\,500, 5\,000, 10\,000$ where $k$ is the position at which we cut-off the ranking, and
    \item the ASR accuracy, by computing \VerboseWER{} (\WER{}) where we boost the top-$k$ (i.e., \WERCut{k}; selected after applying the feedback loop described in \S\ref{section:setup:lm}) entity names using the approach described in Section~\S\ref{section:methodology:asr_boosting} while performing ASR on the automatically-synthesized (using a state-of-the-art TTS system; \S\ref{section:setup:tts}) names of trending entities in our test set (\S\ref{section:setup:data}). Hyperparameter $k$ is our budget of entity names we are able to boost at runtime; a value of $k$ beyond $10\,000$ could lead to increased latency in handling user requests.
\end{enumerate*}
Statistical significance of observed differences in per-utterance \WER{} is determined using a two-tailed paired Student's t-test (\Significant{} $p < 0.01$).

\section{Results and discussion}
\label{section:results}

\newcommand{\RQRef}[1]{\textbf{RQ#1}}

In this section, we answer the three research questions posed in \S\ref{section:introduction}.

\begin{table*}[ht]
\nprounddigits{2}%
\caption{Overview of experimental results (all metrics are in percentages). We generate a ranked list of entities according to predicted popularity and boost the top-$k$ ($k = 2\,500, \ldots$) entity names at recognition time. Machine-learned models that use historical entity query interaction statistics can significantly (\Significant{}; \S\ref{section:setup:evaluation}) reduce the \WER{} (best in \textbf{bold}) for trending entities.\label{tbl:results}}%
\newcommand{\Method}[1]{%
\IfStrEqCase{#1}{%
    {none}{\BaselineNone{}}%
    {random}{\BaselineRandom{}}%
    {most_popular_last_week}{\BaselinePopularLastWeek{}}%
    {popular_previous_week_but_previously_unpopular}{\BaselinePopularLastWeekPreviouslyUnpopular{}}%
    {trending_previous_week}{\BaselineTrendingLastWeek{}}%
    {adaboost}{\Adaboost{}}%
    {nn_128_128}{\NN{}}%
    }%
    [unknown]%
}%
\newcommand{\Heading}[1]{%
\IfStrEqCase{#1}{%
    {0}{\textbf{Heuristics}\\}%
    {1}{\textbf{Machine-learned}\\}%
    }%
    [unknown]%
}%
\newcommand{\MethodResultSuffix}[3]{%
\IfBeginWith{#2}{wer}{%
    \IfStrEqCase{#1}{%
        {4}{\Significant{}}%
        {5}{\Significant{}}%
        }%
        [\phantom{\Significant{}}]%
    }{}%
}%
\newcommand{\MethodResultValue}[4]{%
\IfBeginWith{#2}{wer}{%
    \IfStrEqCase{#3}{%
        {2500}{%
            \IfStrEqCase{#1}{%
                {4}{\textbf{#4}}%
                }%
                [#4]}%
        {5000}{%
            \IfStrEqCase{#1}{%
                {4}{\textbf{#4}}%
                }%
                [#4]}%
        {10000}{%
            \IfStrEqCase{#1}{%
                {5}{\textbf{#4}}%
                }%
                [#4]}%
        }%
        [#4]%
    }{#4}%
}%
\centering%
\renewcommand{\arraystretch}{0.85}%
\begin{tabularx}{0.90\textwidth}{lcccccccccc}%
\toprule%
&&\multicolumn{3}{c}{2\,500}&\multicolumn{3}{c}{5\,000}&\multicolumn{3}{c}{10\,000}\\%
&AP&Recall&Precision&WER&Recall&Precision&WER&Recall&Precision&WER\\%
\cmidrule(lr{.50em}){2-2}%
\cmidrule(lr{.50em}){3-5}%
\cmidrule(lr{.50em}){6-8}%
\cmidrule(lr{.50em}){9-11}%
\Method{none}&{-}&{-}&{-}&\MethodResultValue{baseline}{wer}{2500}{\numprint{14.67}\MethodResultSuffix{baseline}{wer}{2500}}&{-}&{-}&\MethodResultValue{baseline}{wer}{5000}{\numprint{14.67}\MethodResultSuffix{baseline}{wer}{5000}}&{-}&{-}&\MethodResultValue{baseline}{wer}{10000}{\numprint{14.67}\MethodResultSuffix{baseline}{wer}{10000}}\\%
\midrule%
\Heading{0}%
\Method{random}&\MethodResultValue{0}{average_precision}{None}{\phantom{0}\numprint{2.5723635926396597}\MethodResultSuffix{0}{average_precision}{None}}&\MethodResultValue{0}{recall_cut_2500}{2500}{\phantom{0}\numprint{1.0030449579079348}\MethodResultSuffix{0}{recall_cut_2500}{2500}}&\MethodResultValue{0}{precision_cut_2500}{2500}{\phantom{0}\numprint{2.2399999999999998}\MethodResultSuffix{0}{precision_cut_2500}{2500}}&\MethodResultValue{0}{wer_cut_2500}{2500}{\numprint{14.5}\MethodResultSuffix{0}{wer_cut_2500}{2500}}&\MethodResultValue{0}{recall_cut_5000}{5000}{\phantom{0}\numprint{2.059824467132366}\MethodResultSuffix{0}{recall_cut_5000}{5000}}&\MethodResultValue{0}{precision_cut_5000}{5000}{\phantom{0}\numprint{2.3}\MethodResultSuffix{0}{precision_cut_5000}{5000}}&\MethodResultValue{0}{wer_cut_5000}{5000}{\numprint{14.5}\MethodResultSuffix{0}{wer_cut_5000}{5000}}&\MethodResultValue{0}{recall_cut_10000}{10000}{\phantom{0}\numprint{4.352498656636217}\MethodResultSuffix{0}{recall_cut_10000}{10000}}&\MethodResultValue{0}{precision_cut_10000}{10000}{\phantom{0}\numprint{2.4299999999999997}\MethodResultSuffix{0}{precision_cut_10000}{10000}}&\MethodResultValue{0}{wer_cut_10000}{10000}{\numprint{14.18}\MethodResultSuffix{0}{wer_cut_10000}{10000}}\\%
\Method{most_popular_last_week}&\MethodResultValue{1}{average_precision}{None}{\phantom{0}\numprint{3.7349463239497385}\MethodResultSuffix{1}{average_precision}{None}}&\MethodResultValue{1}{recall_cut_2500}{2500}{\phantom{0}\numprint{0.16120365394948952}\MethodResultSuffix{1}{recall_cut_2500}{2500}}&\MethodResultValue{1}{precision_cut_2500}{2500}{\phantom{0}\numprint{0.36}\MethodResultSuffix{1}{precision_cut_2500}{2500}}&\MethodResultValue{1}{wer_cut_2500}{2500}{\numprint{14.86}\MethodResultSuffix{1}{wer_cut_2500}{2500}}&\MethodResultValue{1}{recall_cut_5000}{5000}{\phantom{0}\numprint{0.37614185921547555}\MethodResultSuffix{1}{recall_cut_5000}{5000}}&\MethodResultValue{1}{precision_cut_5000}{5000}{\phantom{0}\numprint{0.42}\MethodResultSuffix{1}{precision_cut_5000}{5000}}&\MethodResultValue{1}{wer_cut_5000}{5000}{\numprint{14.95}\MethodResultSuffix{1}{wer_cut_5000}{5000}}&\MethodResultValue{1}{recall_cut_10000}{10000}{\phantom{0}\numprint{0.967221923696937}\MethodResultSuffix{1}{recall_cut_10000}{10000}}&\MethodResultValue{1}{precision_cut_10000}{10000}{\phantom{0}\numprint{0.54}\MethodResultSuffix{1}{precision_cut_10000}{10000}}&\MethodResultValue{1}{wer_cut_10000}{10000}{\numprint{14.58}\MethodResultSuffix{1}{wer_cut_10000}{10000}}\\%
\Method{popular_previous_week_but_previously_unpopular}&\MethodResultValue{2}{average_precision}{None}{\phantom{0}\numprint{2.795264822943968}\MethodResultSuffix{2}{average_precision}{None}}&\MethodResultValue{2}{recall_cut_2500}{2500}{\phantom{0}\numprint{0.16120365394948952}\MethodResultSuffix{2}{recall_cut_2500}{2500}}&\MethodResultValue{2}{precision_cut_2500}{2500}{\phantom{0}\numprint{0.36}\MethodResultSuffix{2}{precision_cut_2500}{2500}}&\MethodResultValue{2}{wer_cut_2500}{2500}{\numprint{14.85}\MethodResultSuffix{2}{wer_cut_2500}{2500}}&\MethodResultValue{2}{recall_cut_5000}{5000}{\phantom{0}\numprint{0.37614185921547555}\MethodResultSuffix{2}{recall_cut_5000}{5000}}&\MethodResultValue{2}{precision_cut_5000}{5000}{\phantom{0}\numprint{0.42}\MethodResultSuffix{2}{precision_cut_5000}{5000}}&\MethodResultValue{2}{wer_cut_5000}{5000}{\numprint{14.97}\MethodResultSuffix{2}{wer_cut_5000}{5000}}&\MethodResultValue{2}{recall_cut_10000}{10000}{\phantom{0}\numprint{0.967221923696937}\MethodResultSuffix{2}{recall_cut_10000}{10000}}&\MethodResultValue{2}{precision_cut_10000}{10000}{\phantom{0}\numprint{0.54}\MethodResultSuffix{2}{precision_cut_10000}{10000}}&\MethodResultValue{2}{wer_cut_10000}{10000}{\numprint{14.64}\MethodResultSuffix{2}{wer_cut_10000}{10000}}\\%
\Method{trending_previous_week}&\MethodResultValue{3}{average_precision}{None}{\phantom{0}\numprint{3.780747433003696}\MethodResultSuffix{3}{average_precision}{None}}&\MethodResultValue{3}{recall_cut_2500}{2500}{\phantom{0}\numprint{0.21493820526598603}\MethodResultSuffix{3}{recall_cut_2500}{2500}}&\MethodResultValue{3}{precision_cut_2500}{2500}{\phantom{0}\numprint{0.48}\MethodResultSuffix{3}{precision_cut_2500}{2500}}&\MethodResultValue{3}{wer_cut_2500}{2500}{\numprint{14.88}\MethodResultSuffix{3}{wer_cut_2500}{2500}}&\MethodResultValue{3}{recall_cut_5000}{5000}{\phantom{0}\numprint{0.6627261329034569}\MethodResultSuffix{3}{recall_cut_5000}{5000}}&\MethodResultValue{3}{precision_cut_5000}{5000}{\phantom{0}\numprint{0.74}\MethodResultSuffix{3}{precision_cut_5000}{5000}}&\MethodResultValue{3}{wer_cut_5000}{5000}{\numprint{14.69}\MethodResultSuffix{3}{wer_cut_5000}{5000}}&\MethodResultValue{3}{recall_cut_10000}{10000}{\phantom{0}\numprint{1.647859573705893}\MethodResultSuffix{3}{recall_cut_10000}{10000}}&\MethodResultValue{3}{precision_cut_10000}{10000}{\phantom{0}\numprint{0.9199999999999999}\MethodResultSuffix{3}{precision_cut_10000}{10000}}&\MethodResultValue{3}{wer_cut_10000}{10000}{\numprint{13.56}\MethodResultSuffix{3}{wer_cut_10000}{10000}}\\%
\midrule%
\Heading{1}%
\Method{adaboost}&\MethodResultValue{4}{average_precision}{None}{\numprint{28.79259834790272}\MethodResultSuffix{4}{average_precision}{None}}&\MethodResultValue{4}{recall_cut_2500}{2500}{\numprint{20.31166039763568}\MethodResultSuffix{4}{recall_cut_2500}{2500}}&\MethodResultValue{4}{precision_cut_2500}{2500}{\numprint{45.36}\MethodResultSuffix{4}{precision_cut_2500}{2500}}&\MethodResultValue{4}{wer_cut_2500}{2500}{\numprint{12.29}\MethodResultSuffix{4}{wer_cut_2500}{2500}}&\MethodResultValue{4}{recall_cut_5000}{5000}{\numprint{33.11839512806735}\MethodResultSuffix{4}{recall_cut_5000}{5000}}&\MethodResultValue{4}{precision_cut_5000}{5000}{\numprint{36.980000000000004}\MethodResultSuffix{4}{precision_cut_5000}{5000}}&\MethodResultValue{4}{wer_cut_5000}{5000}{\numprint{11.74}\MethodResultSuffix{4}{wer_cut_5000}{5000}}&\MethodResultValue{4}{recall_cut_10000}{10000}{\numprint{49.023822317750316}\MethodResultSuffix{4}{recall_cut_10000}{10000}}&\MethodResultValue{4}{precision_cut_10000}{10000}{\numprint{27.37}\MethodResultSuffix{4}{precision_cut_10000}{10000}}&\MethodResultValue{4}{wer_cut_10000}{10000}{\numprint{11.75}\MethodResultSuffix{4}{wer_cut_10000}{10000}}\\%
\Method{nn_128_128}&\MethodResultValue{5}{average_precision}{None}{\numprint{29.023565795908567}\MethodResultSuffix{5}{average_precision}{None}}&\MethodResultValue{5}{recall_cut_2500}{2500}{\numprint{20.849005910800646}\MethodResultSuffix{5}{recall_cut_2500}{2500}}&\MethodResultValue{5}{precision_cut_2500}{2500}{\numprint{46.56}\MethodResultSuffix{5}{precision_cut_2500}{2500}}&\MethodResultValue{5}{wer_cut_2500}{2500}{\numprint{12.3}\MethodResultSuffix{5}{wer_cut_2500}{2500}}&\MethodResultValue{5}{recall_cut_5000}{5000}{\numprint{32.401934443847395}\MethodResultSuffix{5}{recall_cut_5000}{5000}}&\MethodResultValue{5}{precision_cut_5000}{5000}{\numprint{36.18}\MethodResultSuffix{5}{precision_cut_5000}{5000}}&\MethodResultValue{5}{wer_cut_5000}{5000}{\numprint{11.8}\MethodResultSuffix{5}{wer_cut_5000}{5000}}&\MethodResultValue{5}{recall_cut_10000}{10000}{\numprint{48.55812287300734}\MethodResultSuffix{5}{recall_cut_10000}{10000}}&\MethodResultValue{5}{precision_cut_10000}{10000}{\numprint{27.11}\MethodResultSuffix{5}{precision_cut_10000}{10000}}&\MethodResultValue{5}{wer_cut_10000}{10000}{\numprint{11.72}\MethodResultSuffix{5}{wer_cut_10000}{10000}}\\%
\bottomrule%
\end{tabularx}
\end{table*}

\noindent%
\RQRef{1}: Table~\ref{tbl:results} shows a comparison of the ASR system without any entity name boosting (\BaselineNone{}), four heuristic methods (\S\ref{section:setup:baselines}) that generate a ranking of entity names to boost, and the machine learning models (\S\ref{section:setup:ml}) trained on the features outlined in Table~\ref{tbl:features} with various rank cut-offs $k$ (\S\ref{section:setup:evaluation}). None of the heuristics (\S\ref{section:setup:baselines}) contribute a meaningful improvement. In fact, most of the heuristics actually degrade \WER{} ($k = 2\,500$) on our test set of spoken entity names (\S\ref{section:setup:data}). For the machine-learned models (\Adaboost{}, \NN{}; \S\ref{section:setup:ml}), we see that both models improve \WER{} significantly between $10\%$ and $20\%$ (depending on the value of $k$) compared to the system without any entity name boosting.
A qualitative investigation of the predictions shows that the heuristics perform poorly as they rank popular entities highly; as opposed to infrequent entities that are emerging. For example, all heuristics (except \BaselineRandom{}) rank the entity name \emph{Robert Mueller} highest. On the other hand, the rankings predicted by the machine-learned models have \emph{Scripture}, \emph{Six Sigma}, and \emph{duchess} amongst their top-ranked entity names. We will revisit this observation when we discuss \RQRef{3}.
Finally, we used the systems in the bottom two rows of Table~\ref{tbl:results} (i.e., \Adaboost{}, \NN{}) to recognize a test set consisting of a uniform random sample of manually-annotated, general virtual assistant queries (as opposed to entity-bearing knowledge queries) and measured only a negligible degradation of \WER{} (less than $0.01\%$ relative; not shown in table). Hence, we believe our approach can be used without negatively impacting the user experience.

\begin{figure}[ht]
\includegraphics[width=\columnwidth]{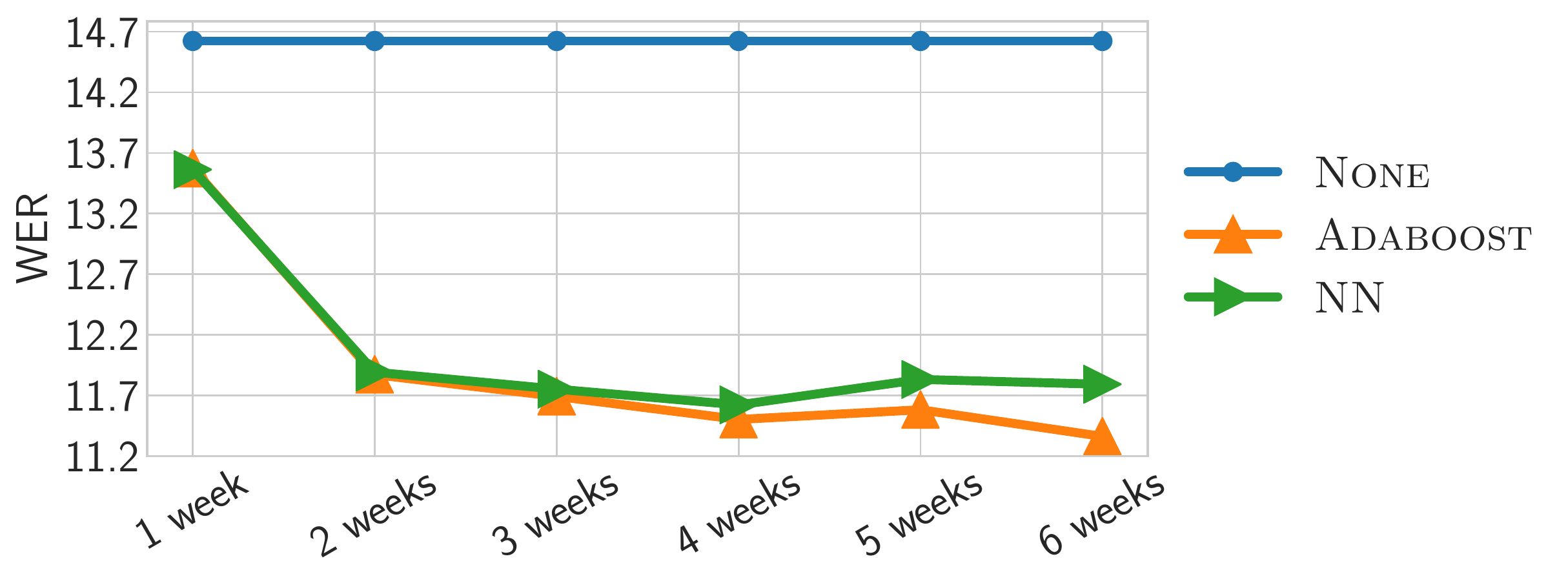}
\caption{The effect of varying the amount of time periods (in weeks) to extract features from; with $k = 5\,000$.\label{fig:sweep}}
\end{figure}

\noindent%
\RQRef{2}: Figure~\ref{fig:sweep} shows the behavior of the \WER{} test metric as we include an increasing amount of history in our models. We can see that even with a single week of history, i.e., models with only feature $\Apply{\Feature{1}}{n - 1}$ as input, we obtain a $7\%$ reduction in \WER{}. Including additional weeks, and hence, rate-of-change features, further reduces \WER{} to around the same level of effectiveness as reported in Table~\ref{tbl:results} (i.e., \numprint{3} weeks of history) and eventually \WER{} reaches a plateau around \numprint{5} to \numprint{6} weeks of history.
Hence, as expected, recent history is more important than far-away history for the entity popularity forecasting task.
However, Figure~\ref{fig:sweep} raises the question regarding why a model with only feature $\Apply{\Feature{1}}{n - 1}$ as input performs better than the heuristic \BaselinePopularLastWeek{} ? We answer this question together with \RQRef{3}.

\begin{figure}[ht]
\includegraphics[width=\columnwidth]{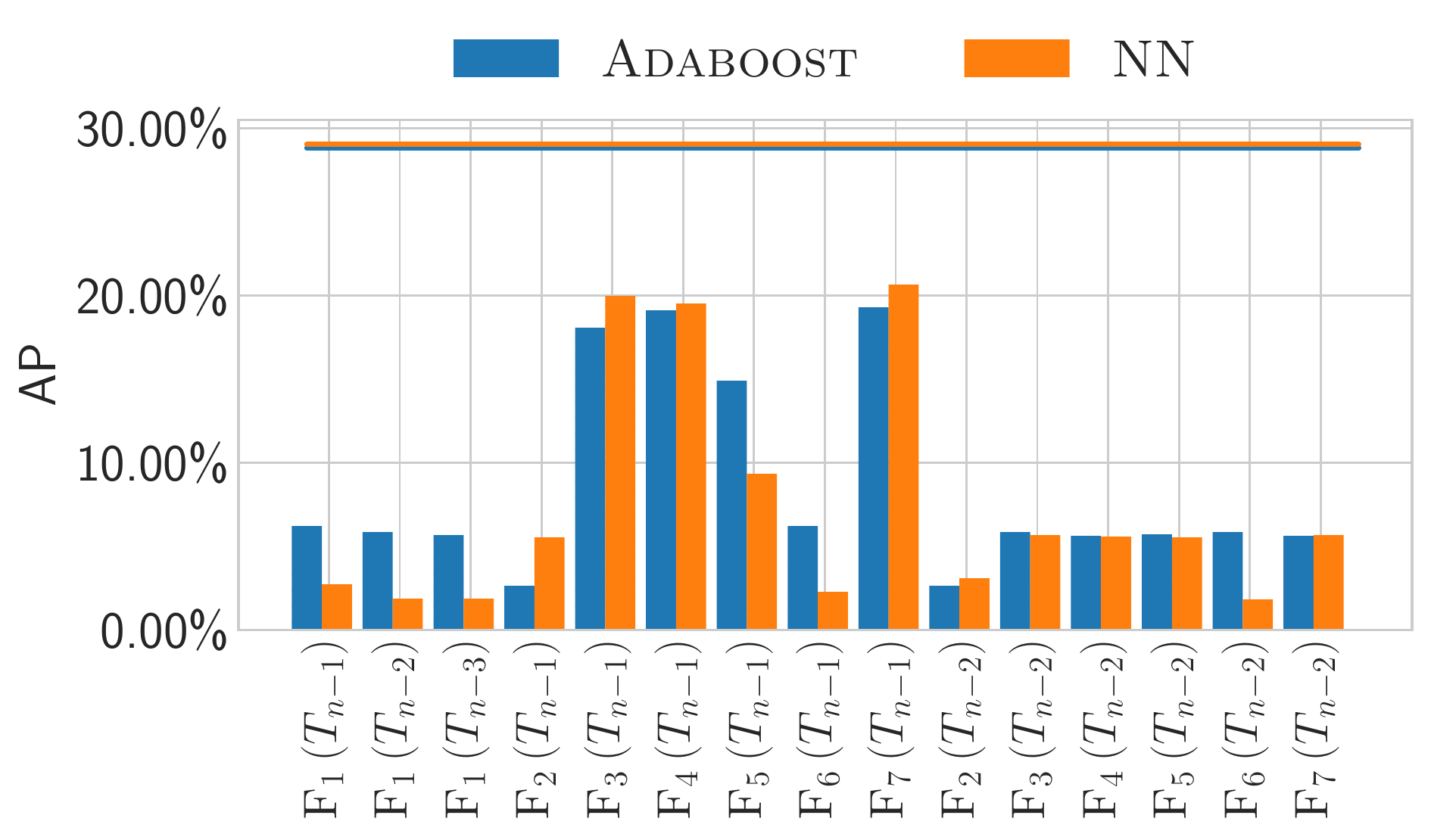}
\caption{Effectiveness of models trained on individual features in terms of \AP{}; with $k = 5\,000$.\label{fig:individual_features}}
\end{figure}

\noindent%
\RQRef{3}: Figure~\ref{fig:individual_features} shows the classification effectiveness (\AP{}) when we train a model for every feature individually ($n = 4$) using our machine learning models (\S\ref{section:setup:ml}). Figure~\ref{fig:individual_features} confirms our findings in \RQRef{2}, namely that rate-of-change features play an important role. However, we note that none of the features individually come close to the same level of \AP{} as was obtained in Table~\ref{tbl:results} by combining all features. In addition, many of our features are correlated (e.g., $\Feature{3}$ is the same as $\Feature{2}$ with the frequencies in log-space), and hence, the effectiveness of individual features is not additive. Consequently, we conclude from Figure~\ref{fig:individual_features} that combining our different features is useful for the entity popularity forecasting task.
Lastly, we revisit the observation raised in \RQRef{1}, where we noted that the heuristics (\S\ref{section:setup:baselines}) are biased towards head entities. Most notably, heuristic \BaselineTrendingLastWeek{} obtains an \AP{} of about $4\%$ in Table~\ref{tbl:results}, whereas the \AP{} of the models trained with \BaselineTrendingLastWeek{} (i.e., $\Apply{\Feature{7}}{\TimePeriod{n - 1}}$) in Figure~\ref{fig:individual_features} is close to $20\%$. The difference here is due to the fact that our machine learning models (\S\ref{section:setup:ml}) learn to threshold feature values (i.e., the decision trees in \Adaboost{} and ReLU activation in \NN{}). Hence, the feature transformations learned by our models are used to filter out head entities and this model property is key to their effectiveness.

\vspace*{-0.25\baselineskip}%
\section{Conclusions}

We introduced a method to improve the speech recognition quality of queries revolving around emerging entities. We trained a model to forecast what entities will become popular from historical entity popularity data, and we subsequently used that information to boost the recognition likelihood of the predicted trending entities.
A comparison of our method with five baselines showed solid improvements of 20\% in word error rate without degrading performance on a test set of general assistant voice queries. Our analyses showed that %
\begin{enumerate*}[label=(\alph*)]
    \item increasing the amount of history---beyond 1 week---used to compute features helps to some extent,
    \item our features are complementary, and
    \item non-linear feature transformations are important for our features to be effective.
\end{enumerate*}
Experimental results not included in this paper show that our findings are consistent over multiple different time periods.
The primary limitation of our work is that we only make use of information contained within the virtual assistant query log. Hence, our approach is limited only to entities that can be recognized already, possibly preceded by additional spoken context that disambiguates the utterance. 
Future work can focus on out-of-vocabulary entities and entities that are often misrecognized, even when additional spoken context is available. In addition, external data sources could provide a useful signal that is not available within the virtual assistant query logs.

\vspace*{-0.25\baselineskip}%
{%
\renewcommand{\acksname}{\textbf{Acknowledgments}}%
\renewcommand{\section}{\oldsubsubsection}%
\begin{acks}
The authors would like to thank Russ Webb, Barry Theobald, Rolf Jagerman, the anonymous, and the internal reviewers for their valuable comments.
\end{acks}
}%

\vspace*{-0.25\baselineskip}%
\bibliographystyle{abbrvnat}%
\small\bibliography{asr-entity-recency}

\end{document}